# Raman micro-spectroscopy as a tool to measure the absorption coefficient and the erosion rate of hydrogenated amorphous carbon films heat-treated under hydrogen bombardment

C. Pardanaud[a], E. Aréou[a], C. Martin[a], R. Ruffe[a], T. Angot[a], P. Roubin[a]
C. Hopf[b], T. Schwarz-Selinger[b], W. Jacob[b]

[a] Aix-Marseille Univ, CNRS, PIIM UMR6633, 13397 Marseille, France

[b] MPI für Plasmaphysik, EURATOM Association, Boltzmannstr. 2, 85748 Garching, Germany.

**Abstract**

We present a fast and simple way to determine the erosion rate and absorption coefficient of hydrogenated amorphous carbon films exposed to a hydrogen atomic source based on ex-situ Raman micro-spectroscopy. Results are compared to ellipsometry measurement. The method is applied to films eroded at different temperatures. A maximum of the erosion rate is found at ~ 450 °C in agreement with previous results. This technique is suitable for future quantitative studies on the erosion of thin carbonaceous films, especially of interest for plasma wall interactions occurring in thermonuclear fusion devices.

## 1. Introduction

Raman micro-spectroscopy is a fast (1 spectrum typically requires ~100 s, with a good signal to noise ratio) and routinely used technique to characterize different forms of C-based materials: diamond, graphite, disordered graphitic materials, amorphous carbons, hydrogenated or not (a-C:H, a-C), etc. The probed surface is a ~ $\mu m^2$ spot and the probed depth ranges from a few tens to a few hundreds of nanometers depending on the sample absorption coefficient. Analysis of the 1000 - 1800 $cm^{-1}$ spectral region gives information on structural (degree of order) and chemical (hybridization of the carbon atoms, H content) properties [1-3]. In the case of an amorphous carbon film deposited on a silicon substrate, measuring the Raman intensity ratio between the main carbon band at ~ 1520 $cm^{-1}$ and the silicon band at ~ 520 $cm^{-1}$ and using Beer's law [4] allow to determine the optical mean free path or the thickness of the film.

Here we use this method to determine the erosion rate and the imaginary part of the refractive index of a-C:H films heat-treated while exposed to an hydrogen atomic source. For these films the thickness and the imaginary part of the refractive index of the as-deposited sample is known. Erosion rates obtained using Raman micro-spectroscopy are compared to those obtained using ellipsometry, the erosion rate of heat-treated a-C:H films varying by more than one order of magnitude between 100 and 450°C [5]. The imaginary part of the refractive index obtained by the two methods can also be compared, the absorption of visible photons significantly increasing with heat-treatments [6]. We finally discuss the opportunity of using Raman micro-spectroscopy for future quantitative studies relevant for magnetic fusion plasma-wall interaction issues.

## 2. Experimental

A hard a-C:H film, with a hydrogen content (H/H+C) of 30 at. % and a density of about 1.7 g $cm^{-3}$ was plasma-deposited at room temperature on a Si [100] wafer (see, e. g., Table 1 in [7], precursor is methane, -300 V bias). The initial thickness ($e_{ad}$ = 193 nm, roughness less than 5 nm) and the imaginary part of the refractive index at the ellipsometry wavelength ($k_{ad}$ = 0.09) were determined by ellipsometry during erosion of an as-deposited sample in an oxygen plasma. Details of the set-up and ellipsometry measurement can be found in [8]. Then this wafer was cut in several pieces and each piece was placed in the oven and exposed to a in-house-built plasma-based atomic hydrogen source under UHV conditions at temperatures ranging between room temperature and 530 °C, and for 30 minutes. Then the H source was switched on between 30 and 600 seconds. For each temperature, the part of the film situated under the mounting brackets was not bombarded (shadowed) and will be referred as to the "non-eroded zone", whereas the rest of the sample is the "eroded zone", i.e. the part of the a-C:H film exposed to the atomic hydrogen source.

Raman spectra were recorded ex-situ on both the non-eroded and the eroded zones of each sample. Raman micro-spectroscopy was performed on a Horiba-Jobin-Yvon HR LabRAM apparatus (laser wavelength: $\lambda_L$ = 514.532 nm, 100X objective, resolution ~ 1 cm$^{-1}$). As our study is based on a comparison of relative band intensities lying at different energies, and as the CCD sensitivity varies with photon energy, we kept the same grating position for all samples analysed, in order to minimize such a bias (which, however, was found to affect the relevant intensity ratios by less than 10 %). We also kept the same optical geometry for all samples analysed in order to minimize biases mentioned in [9]. The laser power was chosen at P ~ 2 or ~ 0.2 mW to have a good signal to noise ratio and/or to prevent damages. A linear background, due to photoluminescence [1], was subtracted. *Ex-situ* ellipsometry measurements were also done for each sample by scanning through the center of the eroded zone starting from the non-eroded zone. In this way the optical constants, assumed to be constant over one sample, as well as the local thickness can be determined.

## 3. Results and discussion

*3.1 Raman spectra*

Figure 1 (a) displays the Raman spectrum of the plasma-deposited hard amorphous carbon film. The intense and narrow band at 520 cm$^{-1}$ together with the weak and broad band at 960 cm$^{-1}$ are due to the Si phonons and combinations. The intense and broad feature in the 1100-1750 cm$^{-1}$ region is composed of two overlapping bands: the D and G bands. These two bands are caused by sp$^2$ carbon atoms: the G band is due to the bond stretching of both aromatic and aliphatic C-C pairs, whereas the D band is due to the breathing of aromatic rings and the presence of disorder [3, 10]. The G band maximum wavenumber, $\sigma_G$, allows the size of the aromatic domains contained in the material to be estimated (local order increasing with this size) [2]. The Raman parameters $\sigma_G$, $I_G$, $I_D$, and $I_{Si}$, where I is the maximum height of the band, were measured directly on the spectra, without any fit. This prevents the problems related to the existence of several solutions to a fitting procedure involving a large number of parameters [11, 12]. Figure 1 (b) displays Raman spectra (1000-1800 cm$^{-1}$ region) recorded in the non-eroded zone of samples heated from room temperature to 530 °C. The D and G bands are always visible. The up-shift of $\sigma_G$ together with the G band narrowing and the increase of $I_D/I_G$ show that the aromatic domain size increases when the sample is heated [10, 13]. The photoluminescence background's slope, usually denoted "m", divided by the G band height is called the "m/$I_G$-parameter". It is known to give an empirical estimate of the H content [1, 14]. Using both these results, we found a value of H/H+C in the range of 24-30 at. % for the as-deposited sample, which is in good agreement with the value determined independently [7]. However thermal heat treatment effects on this parameter will be studied in another publication.

*3.2 Modelling the imaginary part of the refractive index and the erosion rate*

To determine the imaginary part of the refractive index of each heat-treated film, we use the changes in Si and G band intensities together with the knowledge of the thickness and the refractive index of the as-deposited sample. To determine the erosion rate, we compare the Si and G band intensities in the non-eroded and eroded zones to get the thickness of the eroded zone. The method assumes that both incident and Raman light obey Beer's law. This method has been validated by [4] using stylus profilometry for as deposited diamond like carbon samples with thicknesses ranging from 10 nm to 2 µm. Note that the a-C:H/Si and the a-C:H/air interface reflection coefficients are small (~ 0.1) [15] and therefore interference effects inside the a-C:H film are neglected. Then, the G and Si 1$^{st}$ order band intensities are given by (see the derivation in [4]):

$$I_G(T, e, \lambda_G) = \int_0^e I_G(T, z, \lambda_G) \, dz = I_G(T, \infty, \lambda_G) \times (1 - \exp(-2\alpha_G e))$$
$$I_{Si}(T, e, \lambda_{Si}) = \int_e^{+\infty} I_{Si}(T, z, \lambda_{Si}) \, dz \approx I_{Si}(T, \infty, \lambda_{Si}) \times \exp(-2\alpha_{Si} e)$$

(1)

where T is the sample temperature under bombardment, z is the depth inside the a-C:H film or inside the Si substrate, e is the thickness of the a-C:H film, the thickness of the Si substrate being assumed infinite. $\alpha_{G,Si} = 4\pi k(\lambda_{G,Si}, T)/\lambda_{G,Si}$ are the absorption coefficients of the a-C:H film and the Si substrate, respectively, for the wavelength $\lambda_G$ = 560 nm ($\lambda_{Si}$ = 528.7 nm) corresponding to the G band (Si band) photon wavelength using a laser with $\lambda_L$ = 514.5 nm, $k(\lambda_{G,Si}, T)$ being the imaginary part of the a-C:H film refractive index for temperature T and for the wavelength $\lambda_G$ or $\lambda_{Si}$. $I_G(T, \infty, \lambda_G)$ and $I_{Si}(T, \infty, \lambda_{Si})$ are the G and Si band intensities for infinite thickness. Assuming that $\lambda_G \approx \lambda_{Si} \approx \lambda_L = \lambda$ leads to:

$$I_G(T,e) \approx I_G(T,\infty) \times (1-\exp(-2\alpha e))$$
$$I_{Si}(T,e) \approx I_{Si}(T,\infty) \times \exp(-2\alpha e)$$

(2)

For each temperature we measure the intensity ratio:

$$\eta(T) = \frac{I_{Si}(T,e)}{I_G(T,e)} \approx \frac{a(T)}{\exp(2\alpha(T)e)-1} \tag{3}$$

where the quantity $a(T) = I_{Si}(T,\infty)/I_G(T,\infty)$ has to be known in order to determine $\alpha(T)$. Then, we use the knowledge of the absorption coefficient for the as-deposited film (subscript ad) and the hypothesis that a(T) do not vary significantly in the temperature range studied. This hypothesis is supported by the fact that the absolute Raman cross section of the G band does not depend on the aromatic domain size for nano-crystalline graphite [16]. The imaginary part of the refractive index can then be estimated by the formula:

$$k_R(T) \approx \frac{\lambda}{8\pi e_{ad}} \text{Ln}\left(1 + \frac{\eta_{ad}}{\eta(T)}\left(\exp\left(8\pi k_{ad}\frac{e_{ad}}{\lambda}\right)-1\right)\right) \tag{4}$$

For the non-eroded zone (e = $e_{ad}$), figure 2 plots $k_R$ as a function of $k_E$, determined by ex-situ ellipsometry. $k_R$ and $k_E$ measurements were done on two independent sets of samples, labeled as 1 and 2 (sample 1 was then exposed to the hydrogen atomic source whereas sample 2 was not). For the two sets of samples, the dependence on temperature draws a line of slope 1.15, which is close slope 1, showing that the two methods are consistent. However, for $k_E > 0.4$ (530°C) our method underestimates k. As k increases due to the heat treatment, the Raman intensity of the Si wafer decreases, leading to larger error bars. This might explain this underestimation. Note that heating of the a-C:H films leads to relaxation of intrinsic mechanical stress and induces an increase of the film thickness. For hard a-C:H films and annealing to 530 °C the thickness increase amounts to ~12% [17]. The value of $k_R$ taking into account this thickness increase with T and using formula (4) is reduced by only ~2% for T<420°C and is considered as negligible. For T=530°C, the value of $k_R$ is reduced to ~18% and cannot be neglected.

A comparison of spectra recorded in the non-eroded and eroded zones at each temperature was then done. Strong differences were noticed concerning the intensity ratio of the silicon and carbon relative intensity bands. The value of η(T) are useful to determine $e_{er}$(T), the thickness of the sample which has been eroded. However, no sensible differences where noticed on the band shapes attributed to carbon. As an illustration, figure 3 shows the G band wave numbers measured for different temperature. The fact that the values for eroded and non-eroded zones are very close indicates that atomic hydrogen does not induce any sensible structural change of the a-C:H film. From that, we can readily assume that k(T) is most probably similar in the non-eroded and eroded zones. The thickness of the eroded zone $e_{er}$ can thus be expressed as:

$$e_{er}(T) \approx \frac{\lambda_L}{8\pi k(T)} Ln\left(1 + \frac{\eta_{n-er}(T)}{\eta_{er}(T)}\left(\exp\left(8\pi k(T)\frac{e_{ad}}{\lambda_L}\right) - 1\right)\right) \quad (5)$$

where $\eta_{n-er}$ and $\eta_{er}$ are the intensity ratios of the Si and G bands in the non-eroded and eroded zones, respectively. The erosion rate, $R_R$, is finally obtained by dividing the eroded thickness $e_{ad}$-$e_{er}$ by the hydrogen atomic source exposure time. Figure 4 displays the comparison of $R_R$ and the erosion rate determined by *ex-situ* ellipsometry measurements, $R_E$ for sample 1. A linear fit of $R_R$ as function of $R_E$ yields a line with a slope of 1.08 which is reasonably close to the anticipated slope of 1, showing that the two methods are consistent. For temperatures below 250°C, the erosion rate is lower than ~ 0.2 nm s$^{-1}$ and could not been measured. Above 250°C, it increases with increasing temperature up to a value of ~ 2 nm s$^{-1}$ at T ~ 450°C and decreases above this temperature.

We propose one possible application of this following technique. Inside thermonuclear fusion devices erosion of plasma facing components (most often designed with graphitic materials) takes place due to intense particle bombardment (D ions or neutrals, flux ~ 10$^{22}$ m$^2$ s$^{-1}$). This erosion depends on the particle kinetic energy and on the surface temperature. It creates carbon and hydrocarbon species that are transported in the device. During transport, growth may occur ([18] and references therein), leading to nanoparticle formation (nano-onions were found in TEXTOR and Tore Supra, [19]), and carbon deposits are formed on the plasma facing component surfaces [20]. These deposits themselves can be re-eroded and re-deposited. The initial structure of the graphitic carbon is lost, replaced or covered by either amorphous [21] or nano-crystalline [19] carbon layers. Retention of hydrogen isotopes occurs in these layers, which will be a major issue for ITER [22]. To better understand and to model the formation of these hydrogenated deposits, it is of primary interest to understand the fundamental mechanisms that control erosion and erosion parameters determined on well-characterized plasma-deposited a-C:H films [7]. Raman spectroscopy was used here to measure the thickness and k of eroded a-C:H films. This allows studying basic processes occurring during annealing and H-induced erosion of a-C:H films. The presented technique utilizes a silicon wafer because of its strong Raman signal. This indeed limits the application of this method to a limited number of cases. For as deposited samples, it is well suited for thicknesses ranging from 10 nm to 2 μm [4]. For heat treated samples, graphitization modifies the absorption coefficient reducing the range of thicknesses: as an example, for T=530°C, the upper limit is found to be ~200 nm.

## 4. Conclusion

We have demonstrated that *ex-situ* Raman micro-spectroscopy can be used as a fast alternative way to measure the imaginary part of the refractive index and the erosion rate of thin and weakly absorbing a-C:H films deposited on Si wafers.

**Acknowledgments.**

We acknowledge the Euratom-CEA association, the Fédération de Recherche FR-FCM, the EFDA European Task Force on Plasma Wall Interactions, and the French agency ANR (ANR-06-BLAN-0008 contract) for financial support.

**Figure captions**

**Figure 1**: Raman spectra obtained for a plasma-deposited hard amorphous carbon film (a-C:H): (a) as-deposited sample, (b) as-deposited (ad) sample and heat-treated samples from 264 to 458 °C.

**Figure 2:** Imaginary part of the refractive index for the a-C:H film of Fig.1 obtained using Raman and ellipsometry measurements.

**Figure 3:** G band wavenumber, $\sigma_G$, measured for the non-eroded and the eroded zones, for different heating temperatures.

**Figure 4:** Erosion rate obtained using Raman and ellipsometry measurements.

Figure 1

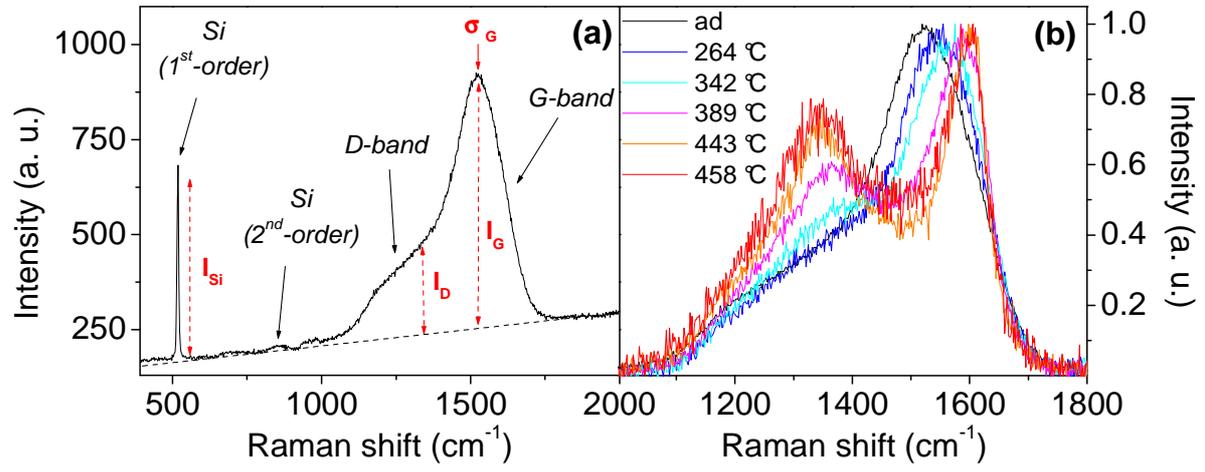

Figure 2

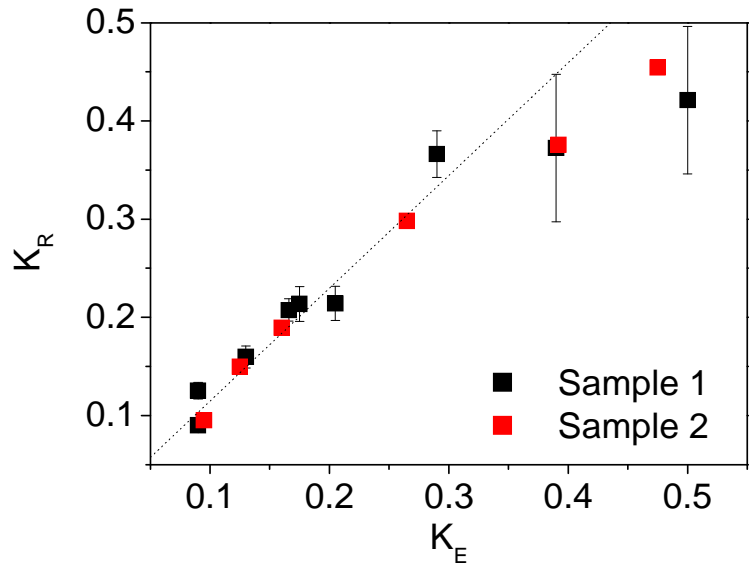

Figure 3

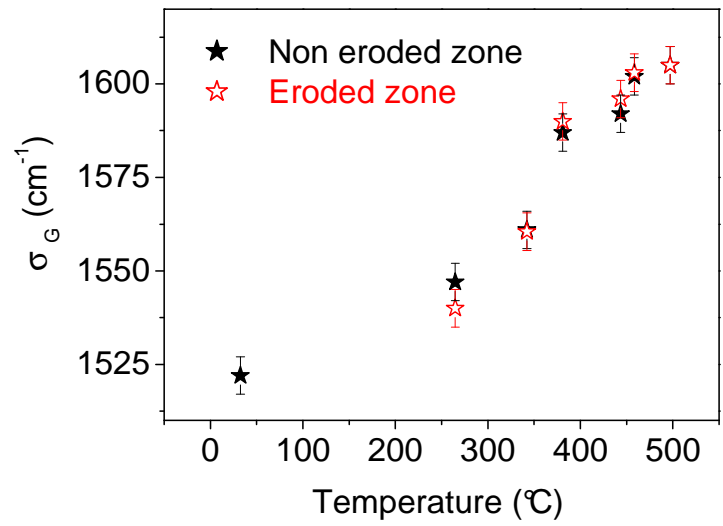

Figure 4

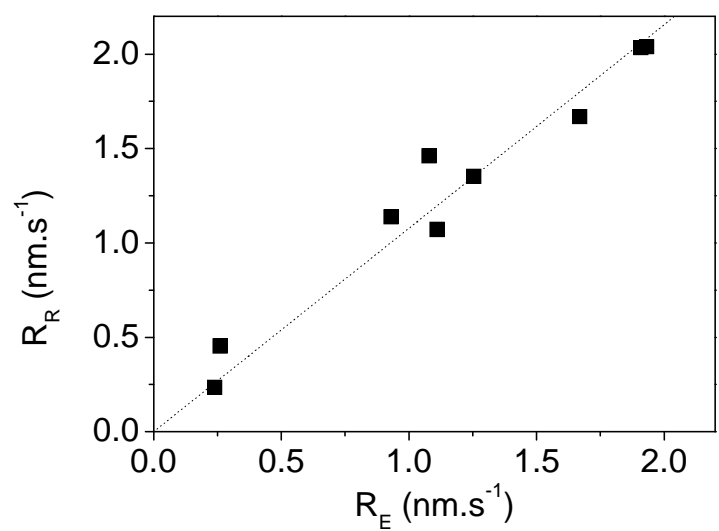